\documentclass[pra,twocolumn,superscriptaddress,longbibliography]{revtex4-1}
\usepackage{amssymb,amsmath,amsfonts,bm}
\usepackage{graphicx}
\usepackage{epstopdf}
\usepackage{times}
\usepackage{float}
\usepackage{lipsum}
\usepackage{color}
\usepackage{ulem}

\usepackage{hyperref}
\hypersetup{colorlinks=true, linkcolor=blue, citecolor=red, urlcolor=blue  }

\usepackage{physics}

\begin{document}

\title{Super-Robust Nonadiabatic Holonomic Quantum Computation in coherence-protected Superconducting Circuits}%

\author{Yuan-Sheng Wang}
 \affiliation{School of Systems Science, Beijing Normal University, Beijing 100875, China}
 \affiliation{School of Electronics and Information Engineering, Suzhou Vocational University, Suzhou 215104, China}

\author{Zhaofeng Su}
 \email{zfsu@ustc.edu.cn}
 \affiliation{School of Computer Science and Technology, University of Science and Technology of China, Hefei 230094, China}
 \affiliation{CAS Key Laboratory of Wireless-Optical Communications, University of Science and Technology of China, Hefei 230026, China}

\author{Xiaosong Chen} \email{chenxs@bnu.edu.cn}
\affiliation{School of Systems Science, Beijing Normal University, Beijing 100875, China} 
\affiliation{Institute for Advanced Study in Physics and School of Physics, Zhejiang University, Hangzhou 310058, China.}
 
\author{Man-Hong Yung}  \email{yung@sustech.edu.cn}

\affiliation{Shenzhen Institute for Quantum Science and Engineering, Southern University of Science and Technology, Shenzhen, 518055, China.}
\affiliation{International Quantum Academy, Shenzhen, 518048, China.}
\affiliation{Guangdong Provincial Key Laboratory of Quantum Science and Engineering, Southern University of Science and Technology, Shenzhen, 518055, China.}
\affiliation{Shenzhen Key Laboratory of Quantum Science and Engineering, Southern University of Science and Technology, Shenzhen, 518055, China.}

\date{\today}

\begin{abstract}
The schmeme of nonadiabatic holonomic quantum computation (NHQC) offers an error-resistant method for implementing quantum gates, capable of mitigating certain errors. However, the conventional NHQC schemes often entail longer operations concerning standard gate operations, making them more vulnerable to the effects of quantum decoherence. In this research, we propose an implementation of the Super-Robust NHQC scheme within the Decoherence-Free Subspace (DFS). 
SR-NHQC has demonstrated robustness against Global Control Errors (GCEs). By utilizing capacitance-coupled transmon qubits within a DFS, our approach enables universal gate operations on a scalable two-dimensional square lattice of superconducting qubits. Numerical simulations demonstrate the practicality of SR-NHQC in DFS, showcasing its superiority in mitigating GCEs and decoherence effects compared to conventional NHQC schemes.  Our work presents a promising strategy for advancing the reliability of quantum computation in real-world applications.
\end{abstract}

                              
\maketitle


\section{Introduction}

Quantum computation is a new paradigm for solving computing tasks based on quantum mechanism~\cite{Nielsen2000}. Due the magic properties of entanglement and nonlocality in quantum mechanics~\cite{HH09,QIP2017,PRA2020}, the corresponding quantum computation has the potential to efficiently solve certain hard problems that are intractable for classical computers, which is known as quantum advantages~\cite{Carmen2019}. The advantage of quantum computation has been not only analyzed in theory but also demonstrated in experiments, such as the simulation of quantum systems\cite{feynman2018simulating,PhysRevLett.112.220501}, factoring of prime numbers \cite{Vandersypen2001,Xu2012,Martin-Lopez2012}, searching unsorted data bases~\cite{Grover1997} and machine learning \cite{Rebentrost2014,PhysRevLett.114.140504,Cong2019}.

However, the current state of quantum computation is in the noisy intermediate-scale quantum (NISQ) era, characterized by the fact that the physical qubits are not enough for building fault-tolerant logical qubits to realize quantum advantage in practical difficult problems~\cite{brooks2019beyond}.
Thus, it is significant to design quantum gates with high-fidelity and robust control. 
Different error suppression protocols have been proposed to mitigate control errors in implementing quantum gates, which include dynamically corrected gates \cite{PhysRevLett.102.080501,PRXQuantum.2.010341}, composite pulses\cite{PhysRevA.70.052318,PhysRevLett.106.240501}, geometric and holonomic quantum computation (GQC and HQC) 
\cite{ZHANG20231}.
Among these protocols, HQC is a promising approach for universally designing robust quantum gates~\cite{liang2023}. 

Quantum holonomies, including Abelian and non-Abelian ones, are global properties of quantum state spaces depending solely on the evolution paths of quantum systems, thereby also referring as the Abelian geometric phases and non-Abelian geometric phases, respectively~\cite{ZHANG20231}. Thus, the logical gates of HQC rely on quantum holonomies inherently possess resilience to a range of quantum errors~\cite{PhysRevLett.91.090404,PhysRevA.72.020301,Leek1889,PhysRevLett.102.030404,PhysRevA.87.060303}. HQC was firstly proposed for adiabatic holonomies \cite{ZANARDI199994} and has been designed for quantum computation based on a variety of physical systems \cite{Duan1695,Wu2005,PhysRevA.76.062311,PhysRevB.83.214518}. 
However, the adiabatic processes are slow in time so that the corresponding applications are susceptible to decoherence and inducing considerable errors. On the contrary, the nonadiabatic approaches are faster and easier to realize than the adiabatic processes.
Therefore, it is more practical to implement quantum gates with nonadiabatic evolutions. And HQC was subsequently generalized to scenarios utilizing nonadiabatic non-Abelian geometric phases \cite{Liu2019,Xugf2012,Xue2015,Xue2017,Ramberg2019,PhysRevResearch.3.L032066}. The nonadiabatic HQC has been experimentally demonstrated in various systems, including superconducting circuits~\cite{AbdumalikovJr2013,Egger2019}, liquid NMR~\cite{Feng2013}, cold atoms~\cite{PhysRevA.78.010302}, trapped ions~\cite{Leibfried2003}, nitrogen-vacancy centers in diamond~\cite{Zu2014,Arroyo-Camejo2014,Sekiguchi2017}, etc.

The author and their collaborators have proposed a new nonadiabatic holonomic quantum computation (NHQC) scheme referred as Super-Robust NHQC (SR-NHQC) in a recent research work~\cite{PhysRevResearch.3.L032066}. 
This scheme provides the enhanced robustness against a specific kind of control error known as the global control error.
The nonadiabatic property allows for a shorter exposure time of qubits to undesired external influences.
However, in most cases, the gate run time of this scenario is still longer than that of standard NHQC and trivial dynamical gates, indicating a higher sensitivity to decoherence.

In this paper, we propose a scheme for implementing the SR-NHQC within a decoherence-free subspace (DFS). This scheme is based on scalable coupling and layout configurations of transmon qubits. We conduct simulations using realistic decoherence parameters to evaluate the performance of various quantum computation protocols, including our proposed method, under the influence of decoherence. By comparing the performance across different schemes, we demonstrate the practicality and superiority of our approach, highlighting its potential for robust and reliable quantum computation.

This paper is organized as follows. In Sec. \ref{sec:general}, we develop universal single-logical-qubit gates in a DFS of three qubits for SR-NHQC. In Sec. \ref{sec:two-qubit}, we introduce a nontrivial two-logical-qubit SR-NHQC gate encoding in a two-excitation subspace of four qubits.  We finally summarize the contribution and outlook of this research in  Sec.~\ref{sec:conclusion}.

\section{\label{sec:general}Universal single-logical qubit gates}
This section introduces the encoding of a single logical qubit within the DFS of three capacitively coupled transmons, and how to implement a universal single-logical-qubit gate. Additionally, we demonstrate how SR-NHQC in DFS effectively mitigates the global control errors (GCEs) in the presence of decoherence.
\subsection{\label{sec:tunable}Tunable coupling through parametric modulations}
For two capacitively coupled transmon qubits $Q_{j_{1}}$ and $Q_{j_{2}}$, the Hamiltonian can be well approximated by \cite{PhysRevApplied.10.054009,PhysRevApplied.10.034050}
\begin{align}
H & =H_{0}+H_{j_{1}j_{2}}, 
\label{eq:h_1q}
\end{align}
in this section, we only consider states in the subspace spanned by $\{|0_{j_{1}}\rangle, |1_{j_{1}}\rangle, |0_{j_{2}}\rangle, |1_{j_{2}}\rangle\}$,
where $|0_{j_{\alpha}}\rangle$ ($|1_{j_{\alpha}}\rangle$) is the ground state (first excited state) of $Q_{j_{\alpha}}$,
in this subspace, the first term of RHS of Eq. \eqref{eq:h_1q} can be written as $H_{0}=\sum_{m=j_{1},j_{2}}\omega_{m}\sigma_{z}^{(m)}/2$ 
 (for simplicity, we set $\hbar = 1$ here and after), 
and the second term reads $H_{j_{1}j_{2}}=g_{j_{1}j_{2}}(\sigma_{j_{1}}^{+}+\sigma_{j_{1}}^{-})(\sigma _{j_{2}}^{+}+\sigma_{j_{2}}^{-})$,
with $\omega_{j}$ the transition frequency of $Q_{j}$, $g_{j_{1}j_{2}} =g_{j_{2}j_{1}}^{*}$ the static coupling strength, $\sigma_{z}^{(j)}$ the Pauli $z$ operator of $Q_{j}$, $\sigma_{j}^{-} =|0\rangle _{j} \langle 1|$ and $\sigma_{j}^{+}=|1\rangle _{j} \langle 0|$ the ladder operators with $\{|0\rangle_{j},|1\rangle_{j}\}$ the computational basis of $Q_{j}$. 

The Hamiltonian in the interaction picture reads $H_{I}(t)=U_{0}^{\dagger}(t)H_{j_{1}j_{2}}(t)U_{0}(t)$, where $U_{0}(t)=\mathcal{T}\exp[-i\int_{0}^{t}H_{0}(t^{\prime})dt^{\prime}]$.
The transmon $Q_{j_{1}}$ can be biased by an AC magnetic flux to periodically modulate its frequency as \cite{PhysRevApplied.10.034050,sciadv.aao3603}: 
\begin{align}
    \omega_{j_{1}}(t)=\omega_{j_{1}}+\epsilon_{j_{1}}\sin(\nu_{j_{1}} t+\phi_{j_{1}}),
\end{align}
by using the Jacobi-Anger identity, $H_{I}(t)$ can be written in the following form
\begin{align}
    H_{I}(t)=&\sum_{n=-\infty}^{\infty}i^{n}g_{j_{1}j_{2}}J_{n}(\beta_{j_{1}})e^{i\Delta_{j_{1}j_{2}}t+i\cdot n(\nu_{j_{1}}t+\phi_{j_{1}})}\sigma_{j_{1}}^{+}\sigma_{j_{2}}^{-} \nonumber\\ 
    &+ \rm{h.c.},
\end{align}
where $\Delta_{j_{1}j_{2}}=\omega_{j_{1}}-\omega_{j_{2}}$, $\beta_{j}=\epsilon_{j}/\nu_{j}$, $J_{n}(\beta)$ is the $n$th Bessel function of the first kind.
When we modulate the driving parameter $\nu_{j_{1}}=-\Delta_{j_{1}j_{2}}$, and then neglecting the high-order oscillating terms, the Hamiltonian in the interaction picture can be approximated as
\begin{align}
H_{I}(t)=g_{j_{1}j_{2}}^{\prime} \sigma_{j_{1}}^{+}\sigma_{j_{2}}^{-}+\text{h.c.},
\label{eq:interaction}
\end{align}
where $g_{j_{1}j_{2}}^{\prime}=g_{j_{1}j_{2}}J_{1}(\beta_{j_{1}})e^{-i(\phi_{j_{1}}+\pi/2)}$. 
We can tune the coupling strength $g_{j_{1}j_{2}}$ by changing $\beta_{j_{1}}$ of the external modulation.
\begin{figure}
    \centering
    \includegraphics[width=0.49\textwidth]{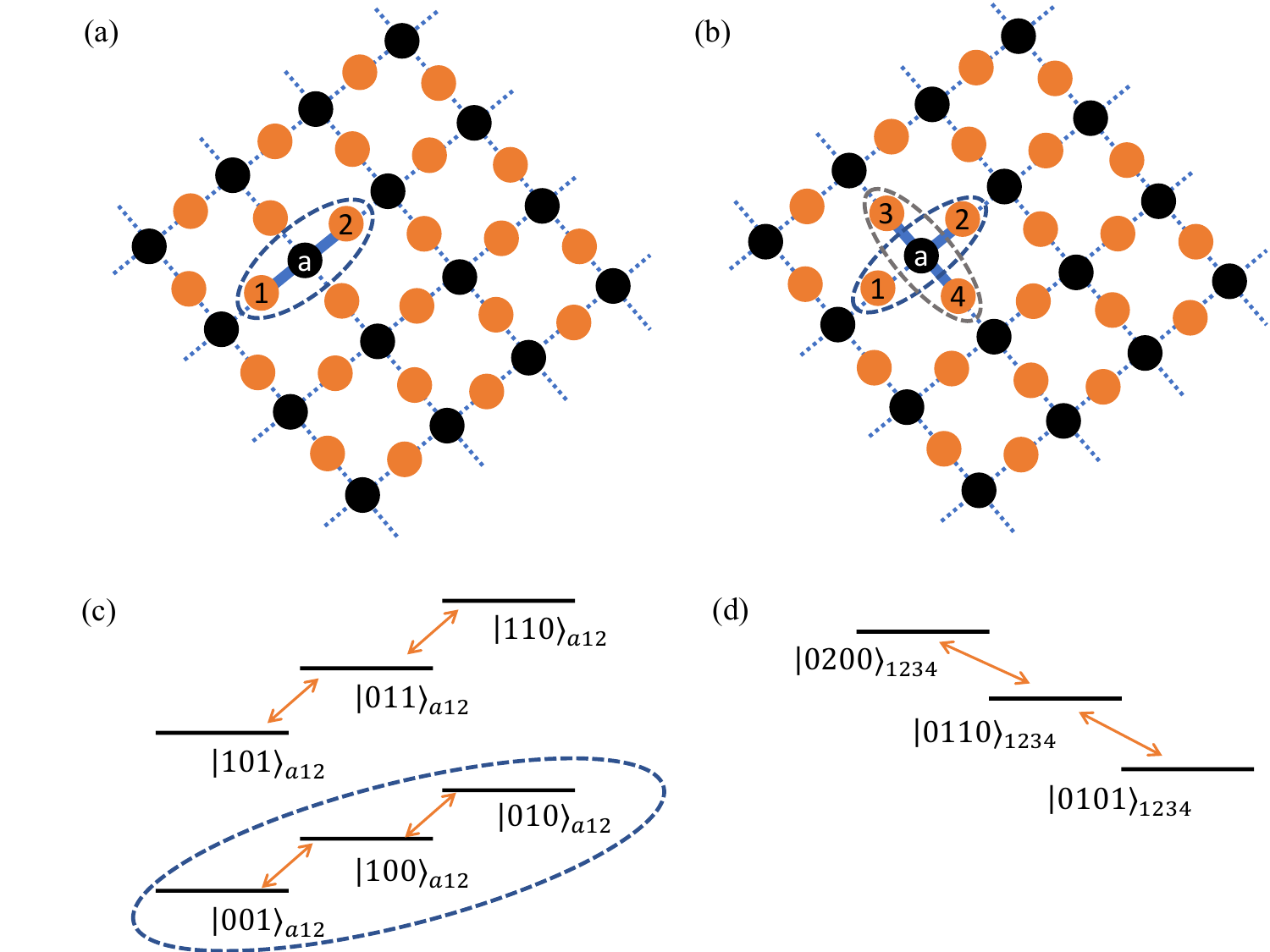}
    \caption{The setup of our proposal: (a) A logical qubit or (b) two coupled logical qubits can be encoded in a scalable square array of coupled superconducting transmons. (c) Energy spectrum for three parametrically tunable coupled qubits, where the single-excitation subspace enables single-logical-qubit holonomic gates. (d) Energy spectrum for four parametrically tunable coupled qubits, where the two-excitation subspace enables the two-logical-qubit holonomic gate.}
    \label{fig:diagram_single}
\end{figure}

\subsection{\label{subsec:single}Conventional NHQC gate in DFS}

The system configuration consists of three transmons, denoted as $Q_{1}$, $Q_{a}$ and $Q_{2}$,
as shown in figure \ref{fig:diagram_single} (a), they are arranged in a linear fashion, with qubit $Q_{a}$ positioned in the middle.
Neighboring transmon qubits are capacitively coupled,
through appropriately biasing $Q_{1}$ and $Q_{2}$ as described in Section \ref{sec:tunable} the Hamiltonian for this three-qubit system in the interaction picture can be approximated as 
\begin{align}
    H(t) = g_{1a}^{\prime}e^{-i\phi_{1}^{\prime}}\sigma_{1}^{+}\sigma_{a}^{-}+g_{2a}^{\prime}e^{-i\phi_{2}^{\prime}}\sigma_{2}^{+}\sigma_{a}^{-}+\text{H.c.}
    \label{eq:h1}
\end{align}
To illustrate the geometric nature of the time evolution operator, we introduce the following dressed states, 
\begin{align}
|b_{1}\rangle &= |0\rangle_{a}\otimes\left[\sin(\frac{\theta}{2})e^{-i\phi}|1\rangle_{L}+\cos(\frac{\theta}{2})|0\rangle_{L}\right], \\  
|b_{2}\rangle &= |1\rangle_{a}\otimes\left[\cos(\frac{\theta}{2})|1\rangle_{L}+\sin(\frac{\theta}{2})e^{i\phi}|0\rangle_{L}\right],
\end{align} 
where $|1\rangle_{L}=|01\rangle_{12}$, $|0\rangle_{L}=|10\rangle_{12}$, $\sin(\theta/2)=g_{2a}^{\prime}/g$, $\cos(\theta/2)=g_{1a}^{\prime}/g$, $g=\sqrt{g_{1a}^{\prime 2}+g_{2a}^{\prime 2}}$, $\phi=\phi_{2}^{\prime}-\phi_{1}^{\prime}$.
Here, $\{|1\rangle_{L},|0\rangle_{L}\}$ are computational basis of the logical qubit,  and $S_{\text{C}}=\text{Span}\{|1\rangle_{L},|0\rangle_{L}\}$ is the corresponding computational subspace.
In terms of dressed states $|b_{1}\rangle$ and $|b_{2}\rangle$, the Hamiltonian \eqref{eq:h1} can be expressed as follows 
\begin{align}
    H(t)=g(t)\left[e^{-i\phi_{1}^{\prime}(t)}|b_{1}\rangle\langle a_1|+e^{i\phi_{1}^{\prime}(t)}|b_{2}\rangle\langle a_2|+\text{H.c.}\right], 
    \label{h_divide}
\end{align}
where $|a_{1}\rangle=|100\rangle_{a12}$, $|a_{2}\rangle=|011\rangle_{a12}$. 
To proceed, we decompose the Hamiltonian $H(t)$ into two parts, namely, $H(t)=H_{1}(t)+H_{2}(t)$, 
where $H_{1}(t)=g(t)e^{-i\phi_{1}^{\prime}(t)}|b_1\rangle\langle a_1|+\text{h.c.}$ and $H_{2}(t)=g(t)e^{i\phi_{1}^{\prime}(t)}|b_2\rangle\langle a_2|+\text{h.c.}$
It is evident that $|a_{1}\rangle$ and $|b_{1}\rangle$ belong to the single-excitation subspace $S_{1}=\rm{Span}\{|001\rangle, |010\rangle, |100\rangle\}$,
while $|a_{2}\rangle$ and $|b_{2}\rangle$ belong to the two-excitation subspace $S_{2}=\rm{Span}\{|011\rangle, |101\rangle, |110\rangle\}$.
Both $S_{1}$ and $S_{2}$ are DFSs since they are insensitive to collective dephasing.
In simpler terms, Hamiltonian \eqref{h_divide} can be seen as the sum of two independent Hamiltonians, each acting on a DFS.
Consequently, collective dephasing does not influence any evolution starting from a state within these DFSs and governed by Hamiltonian \eqref{h_divide}.
Given the independence between $H_{1}$ and $H_{2}$, the temporal evolution operator takes the following form
\begin{align}
    U(t,0)=U_{1}(t,0)\otimes U_{2}(t,0),
\end{align}
where $U_{i}(t,0)=\mathcal{T}\exp(-i\int_{0}^{t}H_{i}(s)ds)$.
For an orthonormal set of basis \(\{|\psi_{k}(0)\rangle\}_{k=1}^{3}\) within the subspace $S_{i}$ ($i\in\{1,2\}$), 
through defining $|\psi_{k}(t)\rangle = U_{i}(t,0) |\psi_{k}(0)\rangle$,
we get another orthonormal basis set: $\{|\psi_{k}(t)\rangle\}_{k=1}^{3}$. Then the evolution operator $U_{i}(t,0)$ can be expressed as
\begin{align}
    U_{i}(t,0) = \sum_{k=1}^{3} |\psi_{k}(t)\rangle \langle \psi_{k}(0)|. 
\end{align}

To proceed, let's consider the third set of basis \(\{|\mu_{k}(t)\rangle\}_{k=1}^{3}\), which span a space denoted as \(M(t)\). When these ancilary basis vectors satisfy the boundary conditions:
\begin{align}
    |\mu_{k}(\tau)\rangle = |\mu_{k}(0)\rangle = |\psi_{k}(0)\rangle,
    \label{eq:bound_cond}
\end{align}
it follows that \(M(\tau) = M(0)\), where \(\tau\) is the total run time of the quantum gate. The continuous variation of \(|\mu_{k}(t)\rangle\) causes \(M(t)\) to move along a smooth and closed path \(\mathcal{C}\) in the \(N\)-dimensional space.
The states \(|\psi_{k}(t)\rangle\) can be expressed in terms of the ancillary basis as $|\psi_{k}(t)\rangle = \sum_{l} C_{lk}(t) |\mu_{l}(t)\rangle$,
where \(C_{lk}(t) = \langle \mu_{l}(t) | \psi_{k}(t) \rangle\). Boundary conditions in Eq.~\eqref{eq:bound_cond} imply that \(C_{lk}(0) = \delta_{lk} C_{ll}(0)\). The temporal evolution operator can then be written as $U(t,0) = \sum_{l,k=1}^{N} C_{lk}(t) |\mu_{l}(t)\rangle \langle \mu_{k}(0)|$.
Substituting this re-expressed \(U(t,0)\) into the Schrödinger equation, we obtain
\begin{align}
\dot{C}_{lk}(t) = \sum_{m=1}^{N} i [A(t) - K(t)]_{lm} C_{mk}(t), 
\label{eq:d_c}
\end{align}
where $A_{lm}(t) = \langle \mu_{l}(t) | i \partial_{t} | \mu_{m}(t) \rangle$ and $K_{lm}(t) = \langle \mu_{l}(t) | H(t) | \mu_{m}(t) \rangle$.
The formal solution of Eq.~\eqref{eq:d_c} is $C_{lk}(t) = \left[ \mathcal{T} e^{i \int_{0}^{t} [A(s) - K(s)] ds} \right]_{lk}$. 

For a conventional NHQC scheme \cite{Sjoeqvist2012}, the Hamiltonian must be carefully designed so that there exists a set of basis vectors \(\{|\mu_{k}(t)\rangle\}_{k=1}^{N}\) satisfying both Eq. \eqref{eq:bound_cond} and the following condition
\begin{align}
    \langle \mu_{l}(t) | H(t) | \mu_{m}(t) \rangle = 0.
    \label{eq:bound_cond_2}
\end{align}
This equation implies $K=0$, then we have
\begin{align}
U(\tau,0) = \sum_{l,k=1}^{N} \left[ \mathcal{T} e^{i \int_{0}^{\tau} A(t) dt} \right]_{lk} |\mu_{l}(0)\rangle \langle \mu_{k}(0)|, 
\end{align}
where $\mathcal{T} e^{i \int_{0}^{\tau} A(t) dt} = \mathcal{P} e^{i \oint_{\mathcal{C}} \mathcal{A}}$,
with $\mathcal{A} = i \langle \mu_{l}(t) | d\mu_{m}(t) \rangle$, and \(\mathcal{P}\) is the path ordering along the closed path \(\mathcal{C}\).
The matrix $\mathcal{T} e^{i \int_{0}^{\tau} A(t) dt}$ depends only on the path \(\mathcal{C}\), not on the detailed form of the system Hamiltonian, thus forming a non-Abelian geometric phase known as holonomy. We may write $U(\tau,0) = U(\mathcal{C})$.
To realize a universal quantum gate, at least two paths \(\mathcal{C}_{1}\) and \(\mathcal{C}_{2}\) are needed, such that \(U(\mathcal{C}_{1})\) and \(U(\mathcal{C}_{2})\) do not commute.
\subsection{\label{sec:robustness}Single-qubit SR-NHQC gate in DFS}
In the SR-NHQC scheme~\cite{PhysRevResearch.3.L032066}, the time evolution operator $U(t,0)$ can be written in the following form
\begin{align}
    U(t,0)=\sum_{k}e^{i\gamma_{k}(t)}|\mu_{k}(t)\rangle\langle\mu_{k}(0)|, \label{eq:u1_t}
\end{align}
i.e. $|\psi_{k}(t)\rangle = e^{i\gamma_{k}(t)}|\mu_{k}(t)\rangle$.
Inserting Eq. \eqref{eq:u1_t} into the Schr\"odinger equation, we have
\begin{align}
    \dot{\gamma}_{k}(t)=\langle\mu_{k}(t)|i\partial_{t}|\mu_{k}(t)\rangle - \langle\mu_{k}(t)|H(t)|\mu_{k}(t)\rangle,
\end{align}
when the parallel transport condition is satisfied, we get
\begin{align}
    \gamma_{k}(\tau) = \int_{0}^{\tau} dt \langle\mu_{k}(t)|i\partial_{t}|\mu_{k}(t)\rangle, \label{eq:g_phase}
\end{align}
$\gamma_{k}(\tau)$ does not depend on the details of the system Hamiltonian and is of geometric origin.

To enable the implementation of universal single-logical-qubit gates in the DFS $S_{1}$, the ancillary states are parametrized as follows:
\begin{eqnarray}
    \begin{split}
    |\mu_{d_{1}}\rangle&=|d_{1}\rangle, \\
    |\mu_{b_{1}}(t)\rangle&=\cos\frac{\Omega(t)}{2}|b_1\rangle-i\sin\frac{\Omega(t)}{2}e^{i\phi_{1}^{\prime}(t)}|a_{1}\rangle, \\
    |\mu_{a_{1}}(t)\rangle&=-i\sin\frac{\Omega(t)}{2}e^{-i\phi_{1}^{\prime}(t)}|b_1\rangle+\cos\frac{\Omega(t)}{2}|a_{1}\rangle.
    \end{split}
    \label{eq:a-states}
\end{eqnarray}
where $|d_{1}\rangle=|0\rangle_{a}\otimes\left[\cos(\frac{\theta}{2})|1\rangle_{L}-\sin(\frac{\theta}{2})e^{i\phi}|0\rangle_{L}\right]$ is a dark state for time evolution governed by $H_{1}$ and $\Omega(t)=\int_{0}^{t}g(s)ds$. 
It is noteworthy that $|\mu_{b1}(0)\rangle=|b_{1}\rangle$, $|\mu_{a1}(0)\rangle=|a_{1}\rangle$. 
Moreover, it can be verified that these ancillary states satisfy the parallel transport condition in Eq.~\eqref{eq:bound_cond_2}. 
When $\Omega(\tau)=n\cdot 2\pi$, with $n$ an integer, $|\mu_{b_{1}}(t)\rangle$ and $|\mu_{a_{1}}(t)\rangle$ are cyclic.

By using the boundary conditions, $U_{1}(\tau,0)$ in the subspace $\text{span}\{|001\rangle,\ |010\rangle\}$ reads $U_{1}(\tau,0)=e^{i\gamma_{b1}(\tau)}|b_{1}(0)\rangle\langle b_{1}(0)|+|d_1\rangle\langle d_1|$.
It can be rewritten as follows 
\begin{align}
    U_{1}(\tau,0)=|0\rangle_{a}\langle 0|\otimes e^{i\gamma_{b1}/2}e^{-i\gamma_{b1}\mathbf{n}_{1}\cdot\bm{\sigma}/2},
    \label{eq:u1}
\end{align}
where $\mathbf{n}=(\sin(-\theta)\cos\phi,\sin(-\theta)\sin\phi,\cos(-\theta))^{T}$ is a unite vector in $\mathbb{R}^{3}$ and $\bm{\sigma}=(\sigma_{x}^{(L)},\sigma_{y}^{(L)},\sigma_{z}^{(L)})^{T}$ are the Pauli operators acting on $S_{C}$ with explicit expressions as follows:
\begin{align}
    \begin{split}
    \sigma_{x}^{(L)}&=|1\rangle_{L}\langle 0|+|0\rangle_{L}\langle 1|, \\
    \sigma_{y}^{(L)}&=-i|1\rangle_{L}\langle 0|+i|0\rangle_{L}\langle 1|,\\
    \sigma_{z}^{(L)}&=|1\rangle_{L}\langle 1|-|0\rangle_{L}\langle 0|.
    \end{split}
\end{align}
Eq. \eqref{eq:u1} meaning that when the initial state of $Q_{a}$ is $|0\rangle_{a}$, we can implement a universal single-qubit gate in the computational subspace $S_{C}$.

In a similar way, for $U_{2}(t,0)$, the following ancillary states are considered 
\begin{eqnarray}
    \begin{split}
    |\mu_{d_{2}}\rangle&=|d_{2}\rangle, \\
    |\mu_{b_{2}}(t)\rangle&=\cos\frac{\Omega(t)}{2}|b_2\rangle-i\sin\frac{\Omega(t)}{2}e^{-i\phi_{1}^{\prime}(t)}|a_{2}\rangle, \\
    |\mu_{a_{2}}(t)\rangle&=-i\sin\frac{\Omega(t)}{2}e^{i\phi_{1}^{\prime}(t)}|b_2\rangle+\cos\frac{\Omega(t)}{2}|a_{2}\rangle,
    \end{split}
    \label{eq:a-states_2}
\end{eqnarray}
where $|d_{2}\rangle=|1\rangle_{a}\otimes\left[\sin(\frac{\theta}{2})|1\rangle_{L}-\cos(\frac{\theta}{2})e^{i\phi}|0\rangle_{L}\right]$ is a dark state.
These ancillary states satisfy the boundary and parallel transport conditions in Eq. \eqref{eq:bound_cond} and Eq. \eqref{eq:bound_cond_2}. 
Then the temporal evolution operator in the subspace spanned by $\{|101\rangle, |110\rangle\}$ can be written as $U_{2}(t,0)=\sum_{k=b_{2},d_{2}}e^{i\gamma_{k}(t)}|\mu_{k}(t)\rangle\langle \mu_{k}(0)|$. 
By using Equations \eqref{eq:a-states_2}, it can be simplified into 
\begin{align}
    U_{2}(\tau,0)=|1\rangle_{a}\langle 1|\otimes e^{i\gamma_{b2}(\tau)/2}e^{i\gamma_{b2}\mathbf{n}_{2}\cdot\bm{\sigma}/2},
    \label{eq:u2}
\end{align}
where $\mathbf{n}_{2}=(\sin\theta\cos\phi,\sin\theta\sin\phi,\cos\theta)^{T}$ is also a unit vector in $\mathbb{R}^{3}$, $\gamma_{k}(\tau)$ have the same form as Eq. \eqref{eq:g_phase} and therefore is a geometric phase. 
Equations \eqref{eq:u1} and \eqref{eq:u2} imply that, conditional on the initial state of the ancillary qubit $Q_{a}$, the temporal evolution operator $U(\tau,0)=U_{1}(\tau,0)\otimes U_{2}(\tau,0)$ corresponds to the rotational gate with rotational axis $\mathbf{n}_{i}$ and rotational angle $\gamma_{bi}$. Therefore, by choosing appropriate $\mathbf{n}_{i}$ and $\gamma_{bi}$, it is possible to realize arbitrary single-qubit gates in the DFS $S_{C}$.

\begin{figure}
    \centering
    \includegraphics[width=0.49\textwidth]{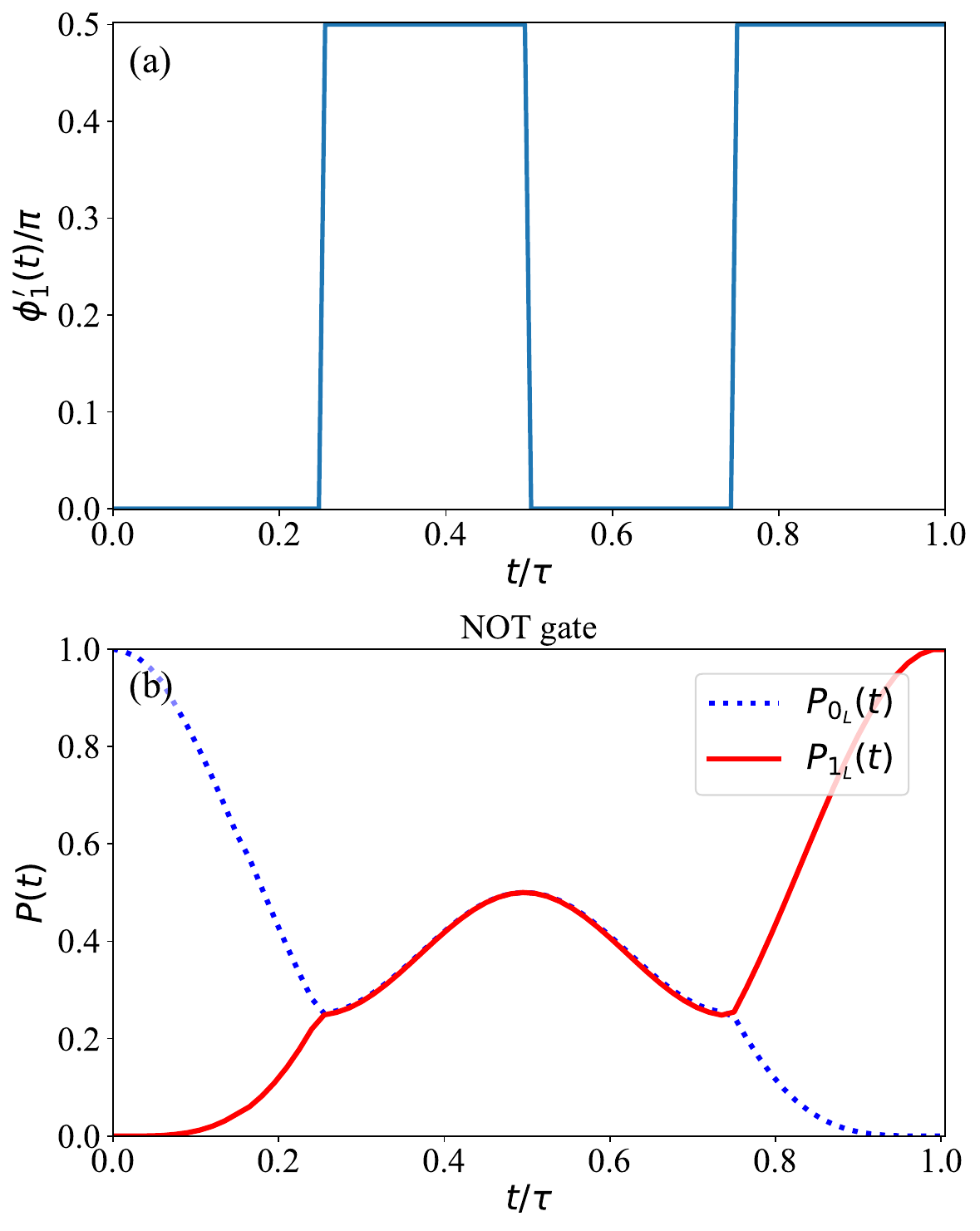}
    \caption{(a) The shape of $\phi_{1}^{\prime}(t)$ for the SR-NHQC NOT gate in DFS. 
    (b) The temporal evolution of populations for states $|0\rangle_{L}$ (blue dotted line), and $|1\rangle_{L}$ (red solid line).}  
    \label{fig:f_v_k1k2}
\end{figure}

To ensure the robustness of SR-NHQC against the GCEs, we have found that in addition to the parallel transport and cyclic boundary conditions,
there is a need for additional constraints to be imposed on the control \cite{PhysRevResearch.3.L032066}. 
These extra constraints are expressed as
\begin{align}
    \int_{0}^{\tau}\langle\psi_{ai}(t)|H_{i}(t)|\psi_{bi}(t)\rangle dt=0,
    \label{r-condition-1}
\end{align}
with $i=1$ and $2$. 
In the subsequent discussion, we set $i=1$ and utilize the NOT gate as an illustrative example to demonstrate the implications of these conditions on the robustness of SR-NHQC gates. 
By substituting the expressions of $H_{1}(t)$, $|\psi_{b1}(t)\rangle$ and $|\psi_{a1}(t)\rangle$ into Eq.\eqref{r-condition-1}, 
we can rewrite the condition as $\int ^{\tau }_{0} dt \exp[i\left( 2\gamma_{b1}( t) +\phi ^{\prime }_{1}( t)\right)] =0$.
In accordance with the relationship $H_{1}(t)=\sum_{\alpha_{1}}i|\dot{\psi}_{\alpha_{1}}(t)\rangle\langle \psi_{\alpha_{1}}(t)|$ and Eq. \eqref{h_divide}, 
we can deduce $\gamma _{b1}(t)=\dot{\phi}_{1}^{\prime}(t)/\cos \Omega(t)$. 
Consequently, the aforementioned condition can be expressed in terms of $\phi_{1}^{\prime}(t)$ and $\Omega(t)$ as:
\begin{align}
 \int ^{\tau }_{0} dt e^{i\int_{0}^{t}\left[\dot{\phi}_{1}^{\prime}(t)/\cos \Omega(t)\right]dt} & =0.    
 \label{eq:r_cond}
\end{align}
It is noteworthy that the solution of Eq. \eqref{eq:r_cond} is not unique, and one possible solution is given by:
\begin{align}
\phi_{1}^{\prime}(t)&=\begin{cases}
0 & 0\leq t\leq\tau/4, \\
\gamma & \tau/4<t\leq\tau/2, \\
0 & \tau/2<t\leq 3\tau/4, \\
\gamma & 3\tau/4<t\leq\tau.
\end{cases}
\label{eq:phi_1}
\end{align}
The shape of $\phi_{1}^{\prime}(t)$ is shown in Fig. \ref{fig:f_v_k1k2} (a). 
Furthermore, by letting $g=2\pi/\tau$, we have $\Omega(\tau)=2\pi$, the cyclic conditions of ancillary states in Eq. \eqref{eq:a-states} and \eqref{eq:a-states_2} are satisfied. 
To perform a NOT gate, the parameters need to be selected as follows: $\theta=\pi/2$, $\phi=0$, and $\gamma=\pi$. 
When the initial state is $|0\rangle_{L}$, the temporal evolution of populations for $|0\rangle_{L}$ and $|1\rangle_{L}$ are presented in Fig. \ref{fig:f_v_k1k2} (b).

\begin{figure}
    \centering
    \includegraphics[width=0.49\textwidth]{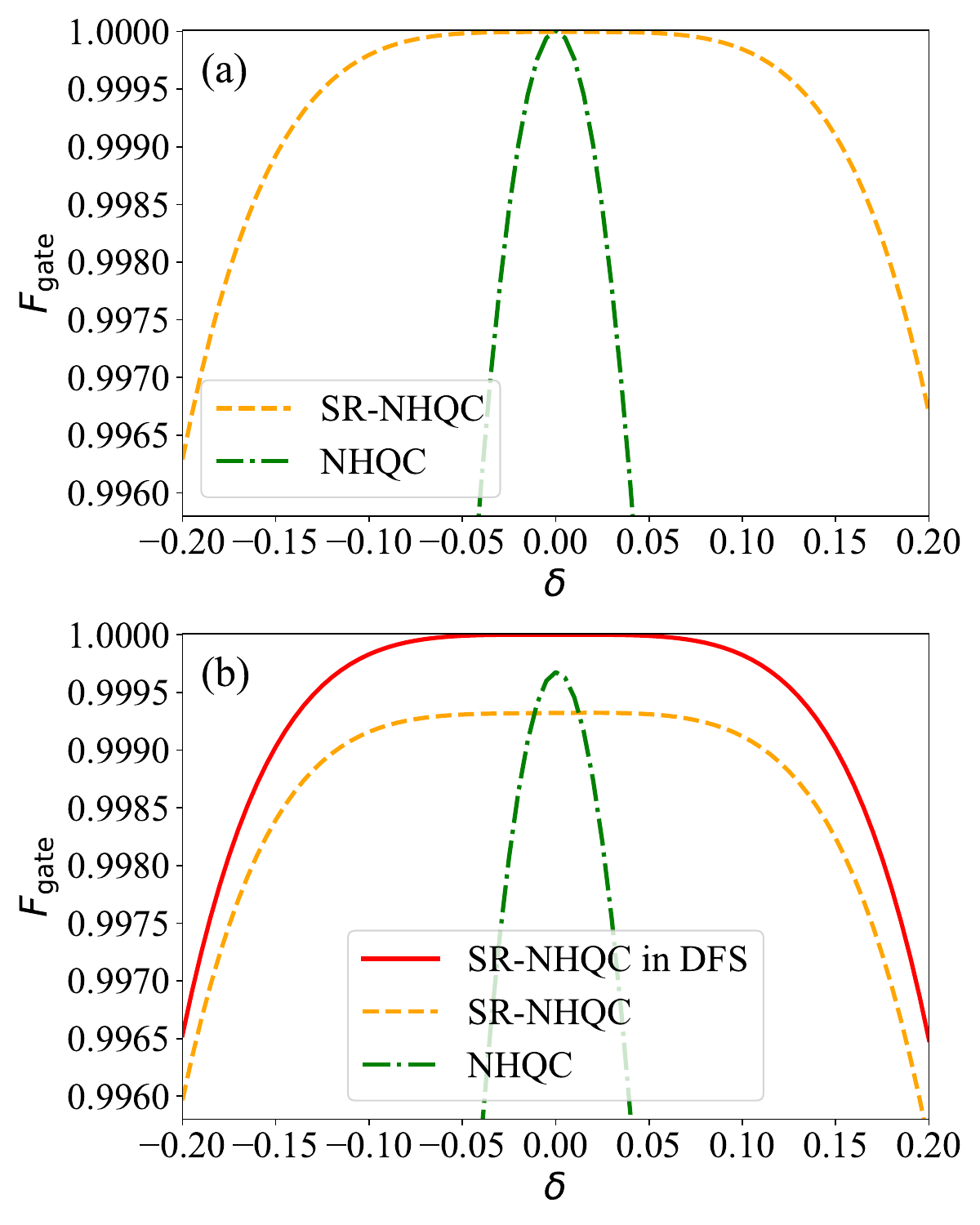}
    \caption{(a) Gate fidelities of the NOT gate versus the amplitude of GCEs in the absence of decoherence. The results of three protocols are shown: SR-NHQC (orange solid line), conventional NHQC (green dashed line), and DG (red dot-dashed line). 
    (b) Gate fidelities of the NOT gate in the presence of collective dephasing. We set $T_{2}=40\ \mu s$ in the simulation. Results of four protocols are shown: SR-NHQC in DFS (blue solid line), SR-NHQC (orange dotted line), NHQC (green dashed line), and DG (red dot-dashed line).
    }
    \label{fig:f_single}
\end{figure}

In the presence of GCEs, the Hamiltonian can be expressed as
\begin{align}
    H^{\prime}(t)=(1+\delta)H(t),
    \label{eq:hp}
\end{align}
where $H(t)$ represents the ideal Hamiltonian and $\delta$ signifies the strength of the error.
Fig. \ref{fig:f_single} (a) depicts how the fidelity of the NOT gate changes with $\delta$ for three scenarios: the standard dynamical gate (DG), the conventional NHQC, and SR-NHQC. 
Compared to the DG and the conventional NHQC scenario, SR-NHQC exhibits superior robustness against GCEs, in the absence of decoherence. 

Apart from control errors, decoherence poses another significant practical challenge in quantum computer development.
Through the numerical simulation, we evaluate and compare the performance of various scenarios considering dephasing and relaxation effects. 
The following master equation is considered to describe the noisy temporal evolution of a transmon's density matrix $\rho(t)$
\begin{align}
    \dot{\rho}(t) = -i[H^{\prime}(t), \rho(t)]+\mathcal{L}[\rho(t)],
    \label{eq:QME_dephasing}
\end{align}
where $H^{\prime}(t)$ is the Hamiltonian given by Eq. \eqref{eq:hp}, $\mathcal{L}[\rho(t)]$ is the Liouvillean given by the following equation
\begin{align}
    \mathcal{L}[\rho(t)] = \frac{\gamma_{2}^{\varphi}}{2}[2\Sigma_{z}\rho(t)\Sigma_{z}-\Sigma_{z}^{2}\rho(t)-\rho(t)\Sigma_{z}^{2}],
    \label{eq:lindblad_operator}
\end{align}
where $\Sigma_{z}=\sum_{j}\sigma_{j}^{z}$ is the collective angular momentum of qubits involved in the gate operations, $\gamma_{2}^{\varphi}$ is the pure collective dephasing rate.

In the numerical simulation, we set the dephasing time $T_{2}=40 \mu$s.  
These values are practical for transmon qubits \cite{annurev-conmatphys-031119-050605}. 
The results are presented in Fig. \ref{fig:f_single} (b). Comparing it to Fig. \ref{fig:f_single} (a), it is evident that although the SR-NHQC scheme demonstrates remarkable robustness to GCE, 
decoherence considerably affects its performance. 
In contrast, the utilization of the decoherence-free subspace effectively mitigates the destructive effects of decoherence,
demonstrated by higher gate fidelities of the SR-NHQC scheme in the DFS.
We also present similar contents for the Hadamard gate in the Appendix \ref{sec:Hgate}. 

\section{\label{sec:two-qubit}Two-logical-qubit gate}
To achieve universal GQC in DFS, a nontrivial two-qubit gate is required besides single-qubit gates.
This section introduces a proposal to perform an SR-NHQC CNOT gate in a DFS.

As illustrated in Fig. \ref{fig:diagram_single} (b), the setup comprises of four transmons, labeled as $Q_{j}$ for $j\in \{1,2,3,4\}$. 
Whin this configuration, $Q_{1}$ and $Q_{2}$ constiute a logical qubit $L_{1}$, while $Q_{3}$ and $Q_{4}$ form another logical qubit $L_{2}$. Notably, $Q_{2}$ is capacitively coupled to both $Q_{3}$ and $Q_{4}$.
The Hamiltonian that describes this system is expressed as follows
\begin{align}
H(t) & =\sum_{j=1}^{4}H_{0,j}+H_{23}+H_{24},
\end{align}
where $H_{0,j}(t)=\sum_{n=1,2}[n\omega_{j}-(n-1)\alpha] |n\rangle_{j} \langle n|$, 
and $H_{ij} =g_{ij}\left( \sigma _{i}^{+} +\sigma _{i}^{-}\right)\left( \sigma _{j}^{+} +\sigma _{j}^{-}\right)$. Here $|n\rangle_{j}$ represents the $(n+1)$th energy level of $Q_{j}$, $\sigma_{j}^{-} =|0\rangle _{j} \langle 1|+\sqrt{2} |1\rangle _{j} \langle 2|$ is a ladder operator, $\alpha$ represents the transmon anharmonicity and $g_{ij}$ maintains the same form as presented in Eq. \eqref{eq:interaction}. 
In contrast to gates involving a single logical qubit, the implementation of a two-logical-qubit gate necessitates the involvement of the second excited state $|2\rangle_{j}$.
We then transform to a rotating frame defined by $H(t)\to U_{0}^{\dagger}(t)[H(t)-i\partial_{t}]U_{0}(t)$, where $U_{0}( t) = U_{0,3} \otimes U_{0,4} \otimes U_{0,2}$,
and $U_{0,j}=\exp\left[-i\int _{0}^{t} dt^{\prime}H_{0,j}(t^{\prime})\right]$.
For $j=1,3,4$, we consider $\omega _{j}(t) =\omega _{j} +\epsilon _{j}\sin( \nu _{j} t+\phi _{j})$, while $\omega _{2}( t) =\omega _{2}$ remains time independent.
After disregarding high-order oscillating terms, the transformed Hamiltonian in the two-excitation subspace can be expressed as
\begin{align}
H & =e^{-i\Phi_{1}(t)} \Big(g^{\prime }_{23} |11\rangle _{23} \langle 20|+g^{\prime }_{24} |11\rangle _{24} \langle 20| \Big)+\text{H.c.}
\label{eq:two-ham}
\end{align}
where $g_{2j}^{\prime}e^{-i\Phi_{1}(t)}=g_{2j}\cdot\sqrt{2}J_{1}(\frac{\epsilon_{j}}{\nu_{j}})e^{-i(\phi_{j}+\pi/2)}$.

In deriving Eq. \eqref{eq:two-ham}, we set $\nu_{j} =\Delta_{2j}-\alpha$, $\phi _{3} =\varphi +\pi /2$, $\phi _{4}=-\pi/2$.
Eq. \eqref{eq:two-ham} can also be reformulated as 
\begin{align}
H &= e^{-i\Phi_{1}(t)} (g^{\prime }_{23} I_{1} \otimes |1\rangle_{2}\langle 2|\otimes |1\rangle_{3}\langle 0|\otimes I_{4} \nonumber \\
&+g^{\prime }_{24} I_{1}\otimes|1\rangle_{2}\langle 2|\otimes I_{3}\otimes|1\rangle_{4}\langle 0|)+\text{H.c.},
\end{align}
where $I_{j} = |0\rangle _{j} \langle 0|+|1\rangle _{j} \langle 1|+|2\rangle _{j} \langle 2|$ represents the identity operator of $Q_{j}$.
Focusing on the two-excitation subspace, the effective Hamiltonian acting on this subspace can be written as 
\begin{align}
H_{L}( t) =G \left( e^{-i\Phi _{1}( t)} |B\rangle \langle A|+e^{i\Phi _{1}( t)} |A\rangle \langle B |\right),
\label{eq:hab}
\end{align}
where $|A\rangle \equiv|0200\rangle _{1234}$, $|B \rangle \equiv \sin(\theta/2) |10\rangle_{L} +\cos(\theta /2)|11\rangle _{L}$,
with $|10\rangle_{L} =|0110\rangle_{1234}$, $|11\rangle_{L}=|0101\rangle _{1234}$,
and $G =\sqrt{(g_{23}^{\prime})^{2} +(g_{24}^{\prime})^{2}}$, $\theta =2\tan^{-1}( g_{23}^{\prime} /g_{24}^{\prime})$.

Similar to the single-logical-qubit case, a two-logical-qubit SR-NHQC gate, there exists a set of ancillary basis $\{\nu_{k}(t)|k=A,B,D\}$ in the subspace spanned by $\{|A\rangle, |10\rangle_{L}, |11\rangle_{L}\}$. This basis states satisfy the following conditions
\begin{enumerate}
    \item[1)] $|\nu_{k}(\tau)\rangle = |\nu_{k}(0)\rangle$, with $\tau$ the run time of the two-qubit gate,
    \item[2)] $|\psi_{k}(t)\rangle = e^{i\gamma_{k}(t)}|\nu_{k}(t)\rangle$, where $\gamma_{k}(t)$ is a real function, $|\psi_{k}(t)\rangle\equiv U(t,0)|\nu_{k}(0)\rangle$, with $U(t,0)$ as the temporal evolution operator, 
    \item[3)] $\langle \nu_{k}(t)|H_{L}(t)|\nu_{k}(t)\rangle = 0$,
    \item[4)] $\int_{0}^{\tau}dt \langle \psi_{k}|H_{L}(t)|\psi_{l}(t)\rangle = 0$ for $k\neq l$.
\end{enumerate}
Here $k,l = A,B,D$. By introducing a dark state defined as $|D \rangle = \cos(\theta /2) |10\rangle_{L} +\sin(\theta/2) e^{-i\varphi} |11\rangle_{L}$, the ancillary states can be parameterized as follows
\begin{align}
    \begin{split}
    |\nu_{D}(t)\rangle&=|D \rangle, \\
    |\nu_{B}(t)\rangle &= \cos\frac{\Omega_{T}(t)}{2}|B\rangle-i\sin\frac{\Omega_{T}(t)}{2}e^{i\Phi_{1}(t)}|A\rangle, \\
    |\nu_{A}(t)\rangle &= -i\sin\frac{\Omega_{T}(t)}{2}e^{-i\Phi_{1}(t)}|B\rangle+\cos\frac{\Omega_{T}(t)}{2}|A\rangle,
    \end{split}
    \label{eq:anc_T}
\end{align}
where $\Omega_{T}(t)=\int_{0}^{t}G(s)ds$ and $\Omega_{T}(0)=0$, leading to $|\nu_{B}(0)\rangle = |B\rangle$, $|\nu_{A}(0)\rangle = |A\rangle$. Condition 1) is satisfied when $\Omega(\tau)$ is an integer multiple of $2\pi$.
Together with Equation \eqref{eq:hab}, condition (3) is automatically satisfied, while condition 4) can be rewritten as follows,
\begin{align}
     \int ^{\tau }_{0} dt e^{i\int_{0}^{t}\left[\dot{\Phi}_{1}(t)/\cos\Omega(t)\right]dt} & =0.    
\end{align}
In the following, we will demonstrate the implementation of a CNOT gate in the DFS $S_{2\text{LQ}}=\text{span}\{|00\rangle_{L}, |01\rangle_{L}, |10\rangle_{L}, |11\rangle_{L}\}$, satisfying conditions 2) and 4), the subscript "2LQ" stands for "two-logical-qubit".
First, we will describe how to perform a NOT gate in the subspace $S_{\text{BD}}\equiv \text{span}\{|B\rangle, |D\rangle\}=\text{span}\{|10\rangle_{L}, |11\rangle_{L}\}$.
Utilizing conditions 1) and 2), the temporal evolution operator at time $\tau$ can be written as: 
\begin{align}
U(\tau,0) =e^{i\gamma_{A}(\tau)} |A\rangle \langle A|+e^{i\gamma_{B}(\tau)} |B\rangle \langle B|+|D\rangle \langle D|,
\label{eq:u_two_logic}
\end{align}
where $\gamma_{A}(\tau)=-\gamma_{B}(\tau)=\int_{0}^{\tau}dt\langle\mu_{A}(t)|i\partial_{t}|\mu_{A}(t)\rangle$ represents a geometric phase.
In the DFS $S_{\text{BD}}$, the temporal evoluation operator can be expressed as:
\begin{align}
U_{L}( \tau ,0)=e^{-i\gamma _{g}/2} e^{i\gamma _{g}\mathbf{n}_{L} \cdot \mathbf{\sigma}_{L} /2},
\label{eq:rot_two}
\end{align}
where $\mathbf{n}_{L} =(\sin \theta \cos \varphi ,\sin \theta \sin \varphi ,-\cos \theta )$ is a unit vector in $\mathbb{R}^{3}$, $\sigma_{L}=(\sigma_{XL},\sigma_{YL},\sigma_{ZL})^{T}$ with $\sigma_{XL}=|11\rangle_{L}\langle 10|+|10\rangle_{L}\langle 11|$, $\sigma_{YL}=-i|11\rangle_{L}\langle 10|+i|10\rangle_{L}\langle 11|$ and $\sigma_{ZL}=|11\rangle_{L}\langle 11|-|10\rangle_{L}\langle 10|$. 
This equation represents a rotation in the subspace $S_{\text{BD}}$ around the axis $\mathbf{n}_{L}$ by the angle $\gamma_{g}$.

To execute a NOT gate, we select $\theta=\pi/2$, $\varphi=0$, $G=2\pi/\tau$. The function $\Phi_{1}(t)$ is defined as follows to simultaneously satisfy conditions 2) and 4):
\begin{align}
\Phi_{1}(t)&=\begin{cases}
0 & 0\leq t\leq\tau/4, \\
\pi & \tau/4<t\leq\tau/2, \\
0 & \tau/2<t\leq 3\tau/4, \\
\pi & 3\tau/4<t\leq\tau.
\end{cases}
\end{align}
The resulting temporal evolution operator is $U_{\text{BD}}=|10\rangle\langle 11|+|11\rangle\langle 10|$. In the extended subspace spanned by $\{|00\rangle, |01\rangle, |10\rangle, |11\rangle\}$, the operator can be written as:
\begin{align}
U_{\text{2LQ}} &= |0 \rangle _{1L} \langle 0|\otimes ( |0\rangle _{2L} \langle 0|+|1\rangle _{2L} \langle 1|) \nonumber \\
 &+|1\rangle _{1L} \langle 1|\otimes ( |0\rangle _{2L} \langle 1|+|1\rangle _{2L} \langle 0|),
\end{align}
which precisely represents a CNOT gate.

\begin{figure}
    \centering
    \includegraphics[width=0.45\textwidth]{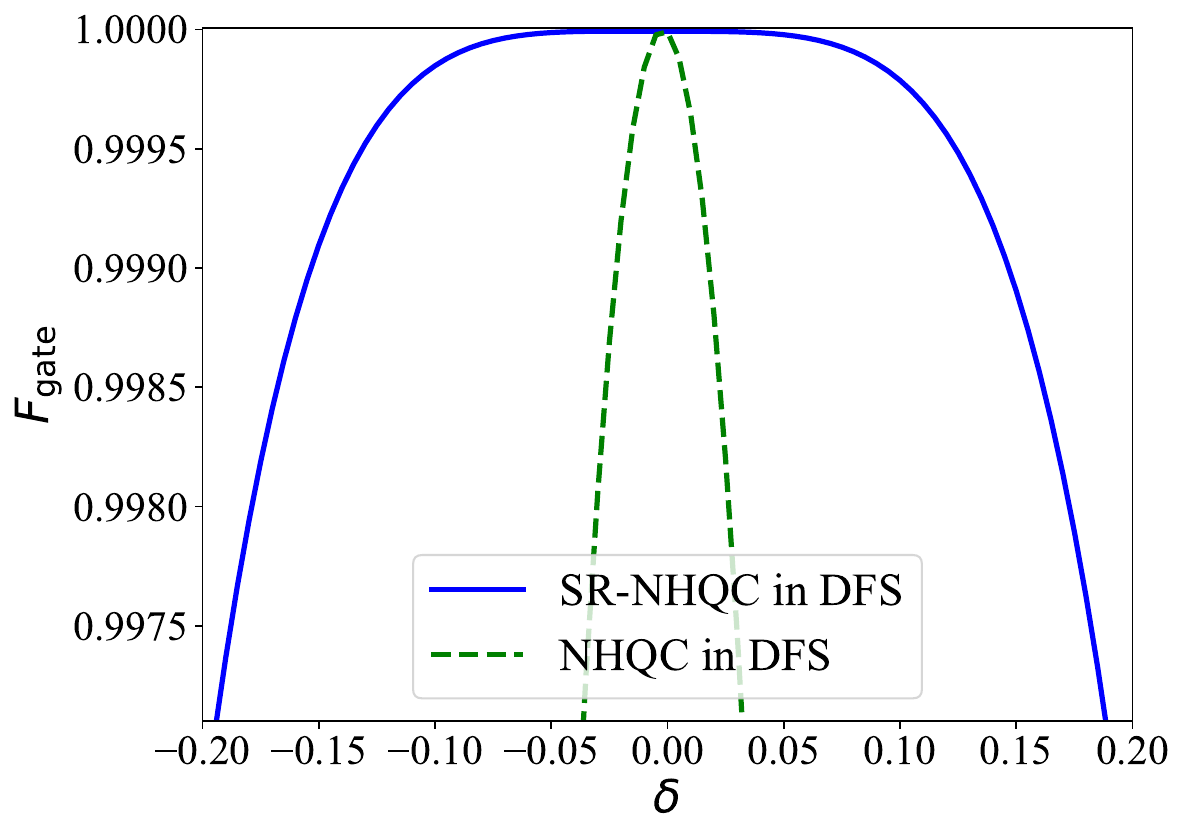}
    \caption{Gate fidelities of the CNOT gate versus the amplitude of GCEs in the presence of decoherence. We set $T_{2}=40\ \mu s$ in the simulation. The results of two protocols are shown: SR-NHQC in DFS (blue solid line) and the standard NHQC in DFS (green dashed line).}
    \label{fig:f_two}
\end{figure}

To assess the robustness of the CNOT gate, we simulated its behavior under the influence of GCEs and collective dephasing, with the dynamics described by the master equation provided in Eqs. \eqref{eq:QME_dephasing} and \eqref{eq:lindblad_operator}. The dephasing time was set to $T_{2} = 40 \mu$s. As illustrated in Fig. \ref{fig:f_two}, the results indicate the superiority of SR-NHQC in suppressing GCEs in the presence of collective dephasing.

\section{\label{sec:conclusion}Conclusion}
In this paper, we introduced a novel approach to address the robustness issues of Nonadiabatic Holonomic Quantum Computation (NHQC) schemes, particularly focusing on the super-robust NHQC (SR-NHQC) framework, which significantly enhances the robustness against global control errors (GCEs). However, while SR-NHQC shows promise in mitigating GCEs, its prolonged operation time compared to conventional NHQC schemes renders it susceptible to decoherence effects, which are ubiquitous in various quantum computing platforms, thereby diminishing its practicality.
To tackle this challenge, we proposed a solution termed SR-NHQC in Decoherence-Free Subspace (DFS) in this work. Our SR-NHQC in DFS approach leverages multiple transmon qubits coupled via capacitance to achieve universal single-qubit gates and a non-trivial two-qubit gate within DFSs. Moreover, the logical qubits in our scheme can be encoded in a scalable two-dimensional square lattice composed of superconducting qubits, a commonly adopted layout for superconducting quantum processors.
To assess the practicality of SR-NHQC in DFS, we conducted numerical simulations evaluating various gate operations, including quantum NOT gates, Hadamard gates, and non-trivial two-qubit gates, under the presence of GCEs and collective dephasing. Comparative analysis confirms the superiority and practicality of our approach in alleviating GCEs compared to alternative schemes.
Our proposed SR-NHQC in DFS scheme presents a promising avenue for enhancing the robustness of NHQC schemes while maintaining practicality in real-world quantum computing implementations. Future research can explore further optimizations and experimental validations to fully harness the potential of this approach in advancing quantum computing technologies.

\begin{acknowledgments}
This research was partially supported by the Innovation Program for Quantum Science and Technology (Grant No. 2021ZD0302901) and National Natural Science Foundation of China (Grants No. 62002333 and No. 12135003).
\end{acknowledgments}

\appendix

\section{Numerical results of the Hadamard gate \label{sec:Hgate}}

In this appendix, we demonstrate the implementation of an SR-NHQC Hadamard gate in a DFS and evaluate its performance in the presence of GCEs and decoherence.

\begin{figure}
    \centering
    \includegraphics[width=0.49\textwidth]{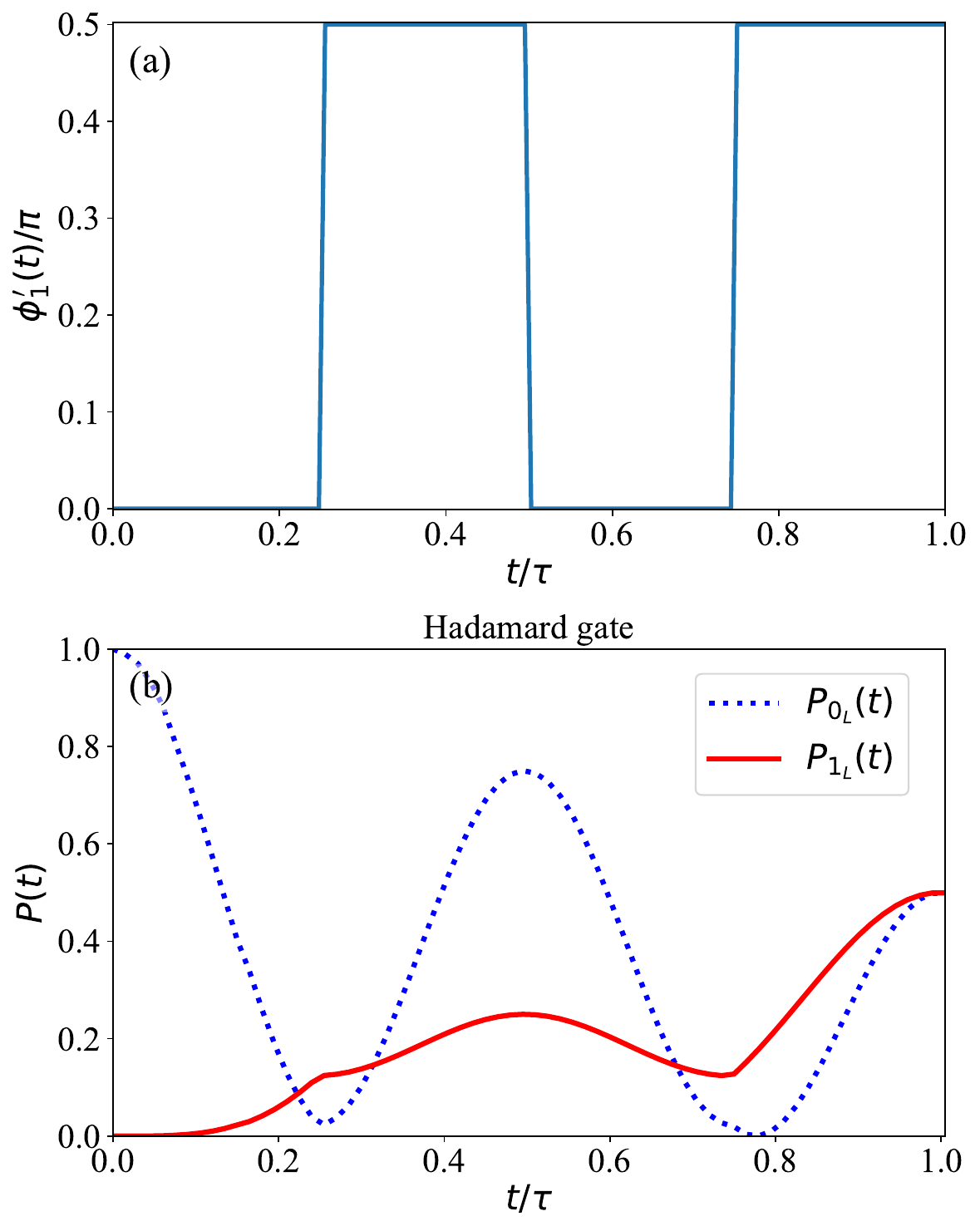}
    \caption{(a) The shape of $\phi_{1}^{\prime}(t)$ for the SR-NHQC Hadamard gate in DFS. 
    (b) The temporal evolution of populations for $|0\rangle_{L}$ (blue dotted line), and $|1\rangle_{L}$ (red solid line).}  
    \label{fig:H_gate_pulse_dyn}
\end{figure}

\begin{figure}
    \centering
    \includegraphics[width=0.49\textwidth]{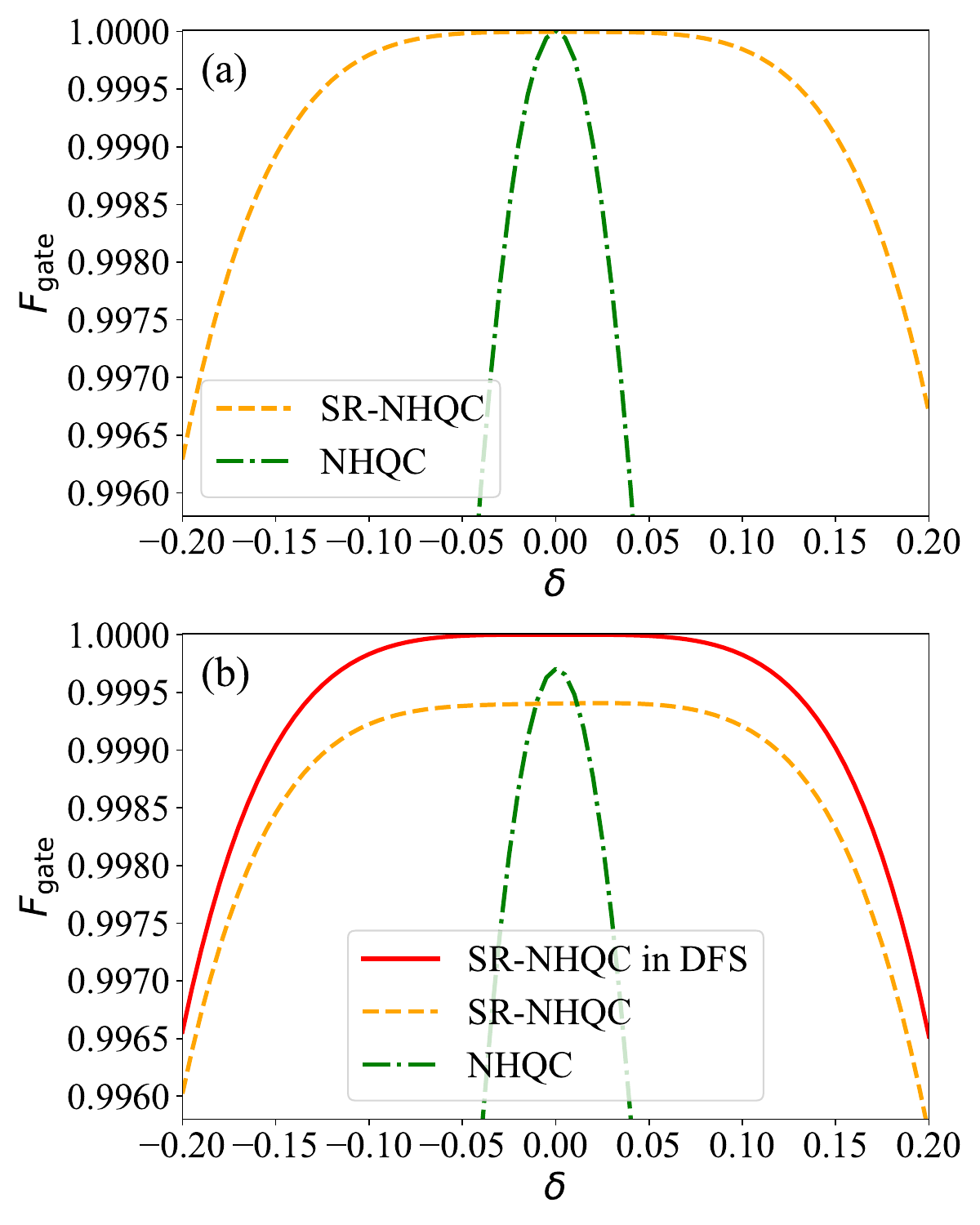}
    \caption{(a) Gate fidelities of the Hadamard gate versus the amplitude of GCEs in the absence of decoherence. The results of three protocols are shown: SR-NHQC (orange solid line), conventional NHQC (green dashed line), and DG (red dot-ashed line). 
    (b) Gate fidelities of the Hadamard gate in the presence of dephasing and relaxation. In the simulation, we set $T_{2}=40\ \mu s$. Results of four protocols are shown: SR-NHQC in DFS (blue solid line), SR-NHQC (orange dotted line), NHQC (green dashed line), and DG (red dot-dashed line).
    }
    \label{fig:H_gate_comb}
\end{figure}

To execute an SR-NHQC Hadamard gate in the DFS, one may set $\theta = \pi/4$, $\phi = 0$. The function $\phi_{1}^{\prime}(t)$ retains the same form as given in Eq. \eqref{eq:phi_1} and its graph is shown in Fig. \ref{fig:H_gate_pulse_dyn} (a).
Starting from the initial state $|0\rangle_{L}$, the temporal evolution of the populations corresponding to $|0\rangle_{L}$ and $|1\rangle_{L}$ are depicted in Fig. \ref{fig:H_gate_pulse_dyn} (b).
In fig. \ref{fig:H_gate_comb} (a), we present how the fidelity of the Hadamard gate varies with the amplitude of GCEs, denoted as $\delta$, for three scenarios: the standard dynamical gate (DG), the standard NHQC, and SR-NHQC. The results exhibit a pattern similar to that of the NOT gate, indicating that SR-NHQC offers enhanced robustness against GCEs in the absence of decoherence.
Numerical evaluations of different scenarios' performance in the presence of collective are shown in Fig. \ref{fig:H_gate_comb} (b). The dephasing time $T_{2}$ in the numerical simulation matches that of the NOT gate described in the main text. It is evident that employing the decoherence-free subspace sigficantly mitigates the detrimental effects of collective dephasing, resulting in higher gate fidelities compared to the original SR-NHQC scheme.

\bibliography{manuscript}

\end{document}